\documentclass{article}

\usepackage{arxiv}

\usepackage[utf8]{inputenc} % allow utf-8 input
\usepackage[T1]{fontenc}    % use 8-bit T1 fonts
\usepackage{hyperref}       % hyperlinks
\usepackage{url}            % simple URL typesetting
\usepackage{booktabs}       % professional-quality tables
\usepackage{amsfonts}       % blackboard math symbols
\usepackage{nicefrac}       % compact symbols for 1/2, etc.
\usepackage{microtype}      % microtypography
\usepackage{lipsum}
\usepackage{graphicx}
\usepackage{textcomp}
\usepackage{longtable}
\usepackage{booktabs}
\usepackage{lscape}
\usepackage[table,xcdraw]{xcolor}
\usepackage{xcolor}
\usepackage{comment}
\usepackage{float}
\title{Automatic Recommendation of Strategies for Minimizing Discomfort in Virtual Environments}

\author{
  Thiago M. Porcino\thanks{} \\
  Institute of Computing\\
  Universidade Federal Fluminense\\
  Niterói, Brazil\\
  \texttt{thiagomp@ic.uff.br} \\
      \And
  Erick O. Rodrigues\\
  Academic Department of Informatics\\
  Universidade Tecnologica Federal do Paraná\\
  Pato Branco, Brazil \\
  \texttt{erodrigues@utfpr.edu.br} \\
       \And
  Alexandre Silva\\
  Departament of Academic Informatics\\
  Instituto Federal do Triangulo Mineiro\\
  Uberaba, Brazil \\
  \texttt{alexandre@iftm.edu.br} \\
    \And
 Daniela Trevisan\\
  Institute of Computing\\
  Universidade Federal Fluminense\\
  Niterói, Brazil\\
  \texttt{daniela@ic.uff.br} \\
   \And
 Esteban Clua\\
  Institute of Computing\\
  Universidade Federal Fluminense\\
  Niterói, Brazil\\
  \texttt{esteban@ic.uff.br} \\
}

\begin{document}
\maketitle

\begin{abstract}
Virtual reality (VR) is an imminent trend in games, education, entertainment, military, and health applications, as the use of head-mounted displays is becoming accessible to the mass market. Virtual reality provides immersive experiences but still does not offer an entirely perfect situation, mainly due to Cybersickness (CS) issues. In this work we first present a detailed review about possible causes of CS. Following,  we propose a novel CS prediction solution. Our system is able to suggest if the user may be entering in the next moments of the application into a illness situation. We use Random Forest classifiers, based on a dataset we have produced.  The CSPQ (Cybersickness Profile Questionnaire) is also proposed, which is used to identify the player's susceptibility to CS and the dataset construction. In addition, we designed two immersive environments for empirical studies where participants are asked to complete the questionnaire and describe (orally) the degree of discomfort during their gaming experience. Our data was achieved through 84 individuals on different days, using VR devices. Our proposal also allows to identify which are the most frequent attributes (causes) in the observed discomfort situations.
\end{abstract}

% keywords can be removed

\section{Introduction}
We are currently experiencing the birth and development of a new entertainment platform. Virtual Reality (VR) delivers immersive 3D graphics in entertainment applications, serious games, and training applications in health, technological, military, or scientific domains.

Meanwhile, most users that experience head-mounted displays activities feel one or more symptoms of sickness, primarily if the user is subjected for a long period of time ~\cite{laffont2017adaptive}. According to Ramsey et al. ~\cite{ramsey1999virtual}, on average, eighty percent of participants who experienced VR with Head Mounted Displays (HMDs) felt discomfort after the first 10 minutes of exposure. Therefore, more extensive VR activities tend to cause stronger discomfort levels. However, the level of discomfort still varies from individual to individual.

\begin{figure*}
\centering
\includegraphics[width=0.9\linewidth]{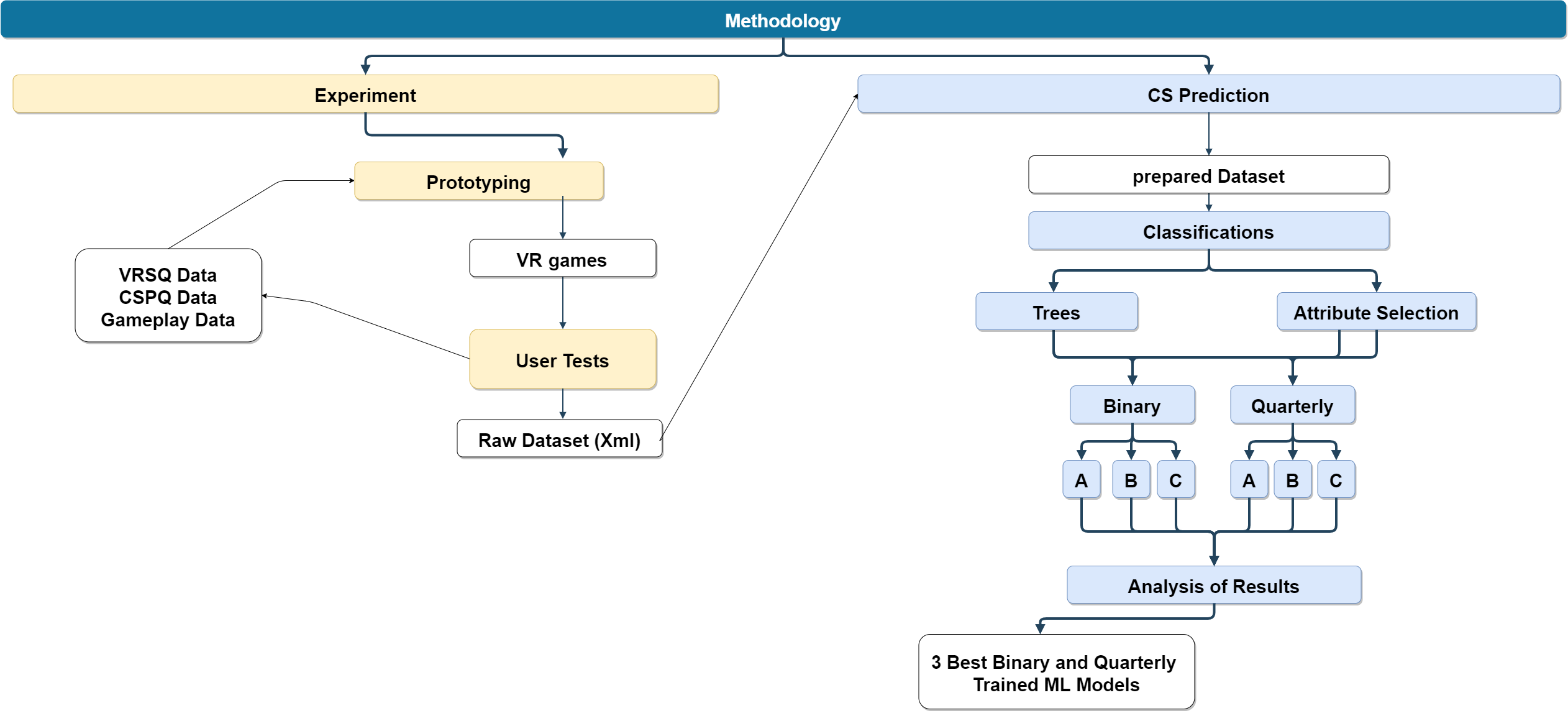}
\caption{The methodology proposed in this work.}
\label{fig:methodology}
\end{figure*}

%\subsection{Motivation}
According to Hua ~\cite{hua20143d}, minimizing sickness in virtual and augmented reality applications is an unresolved challenge. Discomfort resulting from VR can arise from three different leading causes: motion sickness, Visually Induced Motion Sickness (VIMS), CyberSickness (CS) or simulator sickness. 

Motion sickness is manifested by the divergence of information emitted by the human sensory system. This happens when conflicts between the sensory organs that define orientation and position in space occur. Motion Sickness (MS) is defined as the manifestation of discomfort during a forced visual movement (without the movement of the body), for example, during airplane flights, boat trips, or even with land vehicles ~\cite{irwin1881pathology}, ~\cite {lawther1988survey}, ~\cite{bles1998motion}, ~\cite{walker2010head}. This uncomfortable experience also occurs in virtual environments and it is called Visually Induced Motion Disease.

VIMS symptoms are similar to MS. However, the difference lays on the fact that there is no physical movement in VIMS or they are extremely limited ~\cite{keshavarz2015vection}. Several studies categorized VIMS as MS symptoms when situations of visual stimulation manifest irrespective of physical movements. Due to that, different studies in different VIMS contexts were renamed to match the set of symptoms according to the environment they manifested.

Merhi et al. ~\cite {merhi2007motion} define the occurrence of VIMS during experiments with video games as game sickness, Brooks et al. ~\cite {brooks2010simulator} define the occurrence of VIMS in simulators as simulator sickness and McCauley and Thomas ~\cite {mccauley1992cybersickness} define VIMS as occurring specifically in virtual reality systems as CS.
The CS, in turn, is comparable to the symptoms of MS occurring in the real world, such as nausea, vertigo, dizziness, stomach problems, and others ~\cite{howarth1997occurrence}. Symptoms of CS mainly occur with head-mounted displays (Oculus Rift, HTC Vive, among others) ~\cite {rebenitsch2015cybersickness}.

%ta faltando referencia
%The manifestation of VIMS ~\cite{hettinger1992visually},  can originate from several causes, some already known in the literature and others not yet understood.
%Some studies ~\cite{o1973motion}, \cite{reason1975motion} show that MS symptoms, although similar to CS symptoms, do not adhere to the same pattern for manifesting discomfort.%teria que explicar melhor aqui, que padrão? %In this thesis, the study and the resolution of problems related exclusively to CS, that is, only related to the manifestation of discomfort in virtual reality environments with HMDs devices.

%tem que dar uma reorganizada no pensamento dessa parte
%For this reason, many types of research involving sickness in HMDs have been done recently in order to minimize sickness in these devices \cite{kim2018virtual}, \cite{ryge2018preliminary}, \cite{buhler2018reducing}. 
%~\cite{budhiraja2017rotation}, ~\cite{porcino2017minimizing},
%In this research, we map the various causes and symptoms of sickness that occurred during the use of HMDs (be it machine or human actions) ~\cite{rebenitsch2015cybersickness}, \cite{zhang2014human}.

This work investigates the causes and solutions to CS while describing each current methodology. We also propose a novel classification solution for predicting the level of discomfort while using head mounted displays. The literature still lacks a comprehensive description of cybersickness and related topics, with most works consisting of empirical observations and reports ~\cite{wang2019vr}, ~\cite{lee2019motion},~\cite{jeong2019motion}.

As shown in Figure \ref{fig:methodology}, we collected data from a personal questionnaire, from the game play experience while the user was emerged in VR. Classifiers were used to predict the discomfort level of the user over the gameplay. The experiments were separated in binary and quaterly classifications, where in the binary case the labels slight, moderate and severe were merged into one: discomfort. The performance of classifiers is analysed and we also provide an attribute ranking analysis before presenting the conclusions.

%We investigate the causes and solutions in literature ~\cite{langbehn2018evaluation},~\cite{norouzi2018assessing}  in order to expose possible methods towards quantifying the cybersickness. 

%We believe that with this based-literature study, it will be possible to construct a machine learning model to predict and quantify discomfort in VR environments. %and suggest comfort techniques minimize CS in VR environments.
\section{Related Work}

In this section, we review related works, which are addressed in 3 categories related to CS: causes, strategies, and prediction.

%ve se ta ok, pf.
%tem que refazer esse parágrafo aqui fazendo um resumo do que é abordado nessa parte de related work (que tipos de trabalhos relacionados são abordados)
%This section describes the causes ...

\subsection{Causes}
%Taking into account that any symptom of MS occurring in a virtual reality environment with an HMD device is considered CS. 

Several factors can cause pain and discomfort when using head-mounted displays ~\cite {carnegie2015reducing}, ~\cite{yaooculus}. Manifestations of CS can lead to more intense symptoms, such as nausea, eye fatigue, neuralgia, and dizziness ~\cite{kennedy1993simulator}. According to the literature ~\cite{so2001effects}, ~\cite {lin2004virtual}, ~\cite{draper2001effects}, ~\cite{kolasinski1995simulator}, it is possible to highlight the main factors that contribute to the manifestation of CS symptoms and the main strategies to minimize it.

\begin{enumerate}
  \item \textbf{Locomotion} - According to Rebenitsch, 2016 ~\cite {rebenitsch2015cybersickness}, locomotion interaction speed, in this case, the movement, is correlated to cybersickness (CS). Experiences where the participant has greater control of his movements and is close to natural movements of the human body tend to manifest less cybersickness.
    
    \item \textbf{Acceleration} -  Visual accelerations without producing the response in the correspondent vestibular organ cause uncomfortable sensations that result in CS symptoms. High accelerations during movements contribute more to CS  ~\cite{laviola2000discussion}, ~\cite {stanney1997cybersickness}.

    %esse field of view tem que explicar melhor, preferencialmente com imagem
    \item \textbf{Field of View} - Decreasing the FOV display can mitigate the simulator disease ~\cite {yaooculus}, ~\cite {draper2001effects}. However, this also decreases the immersion level.
    
    \item \textbf{Depth of Field} - Inadequate simulation of focus on stereoscopic HMDs with flow tracking devices creates incredible images and, consequently, causes discomfort. In the human eye, focus forces blur effects naturally and according to the depth of field (DoF) and distance range of objects in the observed area ~\cite {zhang2014human}. Due to ocular convergence, objects outside this range, located behind or in front of the eyes, are blurred ~\cite{porcino2017minimizing}.
    
    \item \textbf{Degree of Control} - According to Stanney and Keneddy ~\cite{stanney1997cybersickness} some facts suggest that CS can be mitigated by using experiences with a high level of control in terms of user movements in the virtual world, the authors use the real example of a car driver being much less susceptible to symptoms of discomfort than other passengers who are in the same vehicle.
    
    \item \textbf{Exposure} - In a previous research ~\cite{porcino2016dynamic}, researchers showed that discomfort levels rise proportionally over time. %Consequently, the application should allow users to interrupt the experience to take a rest and then be able to return. %On the other hand, a virtual reality application can recommend users to stop and rest regularly to get around and avoid symptoms of discomfort.  
    
    \item \textbf{Latency} - or lag, has persisted for years as an obstacle in the preceding generations of HMDs ~\cite{olano1995combatting}, ~\cite{oculusblog}. Latency is the time difference between the user input and the correspondent action within a virtual scenario to take place.
    
    \item \textbf{Static Rest Frame} - The lack of a static frame of reference (static rest frame) can cause sensory conflicts and, ultimately, cybersickness.  %nao entendi uma frase que tava aqui (tava sem sentido) entao retirei
    %According to Cao et al. ~\cite{cao2018visually} most users are able to tolerate projection-based VR experiences better (such as CAVEs ~\cite{cruz1992cave}) in relation to HMDs devices. %o que é projection-based vr? referencia
    
    \item \textbf{Camera Rotation} - Rotations in virtual environments with HMDs increase the chances of sensory conflicts occurring. According to the study of Bonato et al. ~\cite{bonato2009combined}, the feeling of vection is greater in rotations when 2 axes are considered in comparison to just 1 axis. The work of Bubka et al. ~\cite{bubka2006rotation} reaffirms the study of Bonato et al. However, Bubka et al. report that many individuals spontaneously reported symptoms of discomfort after many hours after the end of the experience with the VR environment.
    
    \item \textbf{Postural Instability} -  Postural instability (Ataxia) is a postural imbalance or lack of coordination ~\cite{laviola2000discussion} ~\cite{davis2014systematic} caused when the body tries to maintain an incorrect posture due to the sensory conflict caused by the virtual environment. In other words, postural instability is the reactive response to information received by the vestibular and visual organs.
\end{enumerate}

\subsection{Strategies}
Several factors can induce the manifestation of CS symptoms during exposure in virtual reality environments. This manifestation can lead to profound malaise effects, such as nausea, eye fatigue, and dizziness. According to the literature, these problems can be mitigated in several ways: %In recent years, researchers and professionals working with the development of virtual reality applications have used several strategies to reduce CS occurring in VR environments ~\cite{cirio2013kinematic},  ~\cite{carnegie2015reducing}, ~\cite {fernandes2016combating}, ~\cite{kemeny2017new}, ~\cite{melo2018presence}. Such strategies can minimize one or more causes of CS. In this chapter, the strategies studied were described bed and grouped concerning the main known causes of CS manifestation.

\begin{itemize}
    \item \textbf{Locomotion Strategies} - Teleportation techniques assist with the mobility problem in virtual reality environments. In this strategy, the user can travel great distances by specifying the destination point of the trip, using a wand or marker at the destination point ~\cite {langbehn2018evaluation}. Another technique is known as the trigger walk, and this uses the concept of a natural walk to reach a particular destination. With each trigger pull, the user moves one step towards the indicated direction ~\cite{sarupuri2017trigger}.
    
    \item \textbf{Acceleration Strategies} -  According to Berthoz et al. ~\cite{berthoz1975perception} it is possible to induce the sensation of movement using haptic feedback. However, according to Pavard et al. ~\cite{pavard1977linear}, the human visual system can adapt to the illusory movement, but cannot addapt to acceleration. According to Plouzeau et al. ~\cite{plouzeau2018using}, it is possible to measure acceleration as a function of CS in RVs using an EDA (electro-dermal activity). According to surveys ~\cite{tran2017subjective}, ~\cite{kim2017measurement}, the more predictable the movement and acceleration of the camera, the less effects of CS. The technique of slow-motion effects provides less sudden movements and a lower rate of acceleration. This effect works best combined with the blur strategy.
    
    \item \textbf{Field of View Strategies} - Vignette is a technique to gradually reduce the field of view in order to reduce uncomfortable sensations in virtual reality environments ~\cite{fernandes2016combating}. In the work of Norouzi et al. ~\cite {norouzi2018assessing}, a variation of this technique where the vignette size and dynamic field of view (FOV) are related to the camera acceleration was applied. The Tunneling ~\cite{teatherviewpoint} strategy is also used, which reduces the size of the user field of view at the exact moment of the motion.
    
    \item \textbf{Depth of Field Strategies} - Some works include a depth of field simulation (DoF) broker with software blur to minimize the problem of vergence and accommodation ~\cite{carnegie2015reducing}, ~\cite {porcino2016dynamic}. The solution presented by Carnegie and Rhee ~\cite {carnegie2015reducing} pointed to the reduction of the discomfort in HMD applications. More clearly, they suggested a GPU-based solution for DoF simulation in these applications. %In an early work, we developed a focus model for virtual objects with a region of interest (ROI) in real-time. The model uses ROI to determine DoF effects in real-time, minimizing discomfort when using HMD. %nao tem referencia e ta confuso, tirei

    \item \textbf{Rotation Movement Strategies} - Several works applied techniques such as the amplification of movements made by the head ~\cite {kopper2011towards}, ~\cite{plouzeau2018using}. Another example, rotation blurring, which is a technique implemented by Budhiraja et al. ~\cite {budhiraja2017rotation}, applies a smooth gaussian blur to the image based on the magnitude value of the accelerating rotation. 
    
    \item \textbf{Exposure Strategies} -  The CS level of participants increases proportionally in relation to the time of exposure to the VR environment ~\cite {porcino2016dynamic}, ~\cite{melo2018presence}. Therefore, users are advised to periodically take a break out of VR in order to avoid symptoms of discomfort.

    \item \textbf{Rest Frames Strategies} - People tolerate longer exposures without feeling discomfort when projections systems are considered (example: CAVES). One of the biggest differences between VR and projection systems is the rest frames. In projection-based systems, the edges of the screen and elements of the real world are visible beyond the screens and act as resting frames ~\cite{bles1998coriolis}. According to Duh et al. ~\cite {duh2001independent}, VR developers should create suggestions for rest frames whenever possible.
\end{itemize}

\begin{table}
\centering
\setlength{\extrarowheight}{0pt}
\addtolength{\extrarowheight}{\aboverulesep}
\addtolength{\extrarowheight}{\belowrulesep}
\setlength{\aboverulesep}{0pt}
\setlength{\belowrulesep}{0pt}
\caption{Strategies to Overcome CS}
\label{tableStrategies}
\begin{tabular}{ll} 
\toprule
\rowcolor[rgb]{1,1,1}  \textbf{Author(s)}                        & \textbf{Strategies}                \\
Langbern (2018)                                                              & Teleporting                        \\
Farmani (2018)                                                               & Tunneling                          \\
Sapuri (2017)                                                                & Motion Walk                        \\
Berthoz (1975)~                                                              & Haptic Feedback                    \\
Plouzeau (2018)                                                              & Changes on acceleration            \\
Kemeny (2017)                                                                & Headlock                           \\
Skopp (2013)                                                                 & Holosphere                         \\
Cirio (2013)                                                                 & Trajectory Visualization           \\
Budhiraja (2017)                                                             & Rotational blur                    \\
Carnegie (2015)~                                                             & DoF Simulation                     \\
Waveren (2016),~                                                             & Async. Time Warping for Latency    \\
Kim (2012)~                                                                  & "Cabin" Static Frame               \\
Kim (2017)\textasciitilde{}                                                  & Slowmotion                         \\
Bolas (2017)                                                                 & Dynamic FoV                        \\
Norouzi (2018)                                                               & Dynamic Vignetting                 \\
\begin{tabular}[c]{@{}l@{}} Plouzeau (2018) \end{tabular} & Amplified Movements                \\
Hillaire (2008)                                                              & Blur Effects                               \\
Dennison (2016)                                                              & Physiological Signals Observation  \\
\bottomrule
\end{tabular}
\end{table}

\subsection{Prediction}
%precisa reescrever esse paragrafo thiago, tá bem confuso

%Padmanaban et al. ~\cite{padmanaban2018towards} designed a VR sickness predictor. Authors do not focus on minimizing sickness based on VR headset estimated evolution. They use the approach to create a dataset, some questionnaires to evaluate the physiological causes of sickness and individual historical elements to get a more precise result from users. They combine two sickness questionnaires: 
%MSSQ and SSQ to find a single sickness value. They measured SSQ scores from various individuals through stereoscopic content. They used Flownet ~\cite{dosovitskiy2015flownet} to calculate optical flow vectors (they calculate optical flow from one frame to the next, which measures pixel speed). 

%esse ta bem melhor que o anterior, mais direto, limpo, etc
Garcia-Agundez et al. ~\cite{garcia2019development} aimed to classify the level of CS. The model follows a combination of bio-signal and game settings. They collected user signals like respiratory and skin conductivity from a total of 66 participants. As a result, they mentioned a classification accuracy of 82\% for binary classification and 56\% for ternary. 

Jin et al. ~\cite{jin2018automatic} separates factors that cause cybersickness in 3 groups: hardware characteristics (VR device settings and features), software characteristics (the content of the VR scenes), and the individual user. The authors used classifiers to predict the level of discomfort. A total of 3 machine learning algorithms (CNN, LSTM-RNN, and SVR) were used. According to the results, the LSTM-RNN was the most viable model for the case.

%confuso, tem que reescrever pra voltar
%Unlike other works that were using non-interactive content (such as 360 videos), the work of Jin et al. ~\cite {jin2018automatic} performed the cybersickness forecast. However, the authors still focused on collecting user data only after the experience and are limited to quantifying cybersickness.

In this work, we classify the level of cybersickness using parameters from an interactive game scene; this brings the works of Garcia-Agundez et al. and Jin et al. closer to our goal. Both aim to classify CS before and/or after the experience. In this work, we consider the entire VR experience, which means: before, during, and after the participation.

\section{Methodology} \label{cap:cap4}
%Our methodology is divided in the following steps:

%\begin {itemize}
%    \item \textbf{Research}: Identification and association of causes and strategies
%    \item \textbf{Experiment}: Application development, tests with participants, data collection and pre-elimination analysis
%    \item \textbf{CS Automatic prediction}: classifiers, training, binary classification, quartely classification, attribute ranking.
%\end {itemize}

%\begin{figure}
%\centering
%\includegraphics[width=0.9\linewidth]{figuras/method.png}
%\caption{
%Diagram of the methodology applied in this research.}
%\label{fig:Metodologia}
%\end{figure}

%\subsection{Research}

%ta ruim esse paragrafo
%During the investigation stage, an in-depth study was carried out in the existing literature to identify the maximum number of causes %linked to the CS manifestation. Numerous factors can contribute to CS manifestation. Through these works, we associated CS causes and %CS minimization strategies (observed in Table ~\ref{tab:tableStrategiesXCauses}).

%Coloca um parágrafo aqui listando todos os artigos que levaram a criar a tabela abaixo:

\begin{table}[]
\centering
\caption{Strategies associated with causes (1 - Locomotion , 2 - Acceleration, 3 - Field of View, 4 - Depth Of Field, 5 - Degree of Control, 6 - Duration use time, 7 - Latency, 8 - Static rest frame,  9- Camera's rotation, 10 - Postural Instability)}
\label{tab:tableStrategiesXCauses}
\begin{tabular}{|
>{\columncolor[HTML]{C0C0C0}}l |l|l|l|l|l|l|l|l|l|l|}
\hline
\textbf{Strategies X Causes} &
  \cellcolor[HTML]{C0C0C0}1 &
  \cellcolor[HTML]{C0C0C0}2 &
  \cellcolor[HTML]{C0C0C0}3 &
  \cellcolor[HTML]{C0C0C0}4 &
  \cellcolor[HTML]{C0C0C0}5 &
  \cellcolor[HTML]{C0C0C0}6 &
  \cellcolor[HTML]{C0C0C0}7 &
  \cellcolor[HTML]{C0C0C0}8 &
  \cellcolor[HTML]{C0C0C0}9 &
  \cellcolor[HTML]{C0C0C0}10 \\ \hline
Teleporting &
  \multicolumn{1}{c|}{x} &
   &
   &
   &
   &
   &
   &
   &
   &
   \\ \hline
Tunneling &
  \multicolumn{1}{c|}{x} &
   &
   &
   &
   &
   &
   &
   &
   &
   \\ \hline
\begin{tabular}[c]{@{}l@{}}Motion\\ Walk\end{tabular} &
  \multicolumn{1}{c|}{x} &
   &
   &
   &
   &
   &
   &
   &
   &
   \\ \hline
\begin{tabular}[c]{@{}l@{}}Haptic\\ Feedback\end{tabular} &
   &
  \multicolumn{1}{c|}{x} &
   &
   &
   &
   &
   &
   &
   &
   \\ \hline
\begin{tabular}[c]{@{}l@{}}Acceleration\\Changes\end{tabular} &
   &
  \multicolumn{1}{c|}{x} &
   &
   &
   &
   &
   &
   &
   &
   \\ \hline
Headlock &
   &
   &
   &
   &
  \multicolumn{1}{c|}{x} &
   &
   &
   &
   &
  \multicolumn{1}{c|}{} \\ \hline
Holosphere &
  \multicolumn{1}{c|}{x} &
   &
   &
   &
   &
   &
   &
   &
   &
   \\ \hline
\begin{tabular}[c]{@{}l@{}}Trajectory\\ Visualization\end{tabular} &
  \multicolumn{1}{c|}{x} &
   &
   &
   &
   &
   &
   &
   &
   &
   \\ \hline
\begin{tabular}[c]{@{}l@{}}Rotational Blur\end{tabular} &
  \multicolumn{1}{c|}{x} &
   &
   &
   &
   &
   &
   &
   &
  \multicolumn{1}{c|}{x} &
   \\ \hline
\begin{tabular}[c]{@{}l@{}}DoF Simulation\end{tabular} &
   &
   &
   &
  x &
   &
   &
   &
   &
   &
   \\ \hline
\begin{tabular}[c]{@{}l@{}}Latency\\Camera Warping\end{tabular} &
   &
   &
   &
   &
   &
   &
  \multicolumn{1}{c|}{x} &
   &
   &
   \\ \hline
\begin{tabular}[c]{@{}l@{}}"Cabin"\\Static Frame\end{tabular} &
   &
   &
   &
   &
   &
   &
   &
  \multicolumn{1}{c|}{x} &
   &
   \\ \hline
Slowmotion &
   &
  \multicolumn{1}{c|}{x} &
   &
   &
   &
   &
   &
   &
  \multicolumn{1}{c|}{x} &
   \\ \hline
\begin{tabular}[c]{@{}l@{}}Dynamic FoV\end{tabular} &
   &
   &
  \multicolumn{1}{c|}{x} &
   &
   &
   &
   &
   &
   &
   \\ \hline
\begin{tabular}[c]{@{}l@{}}Dynamic\\Vignetting\end{tabular} &
  \multicolumn{1}{c|}{x} &
   &
  \multicolumn{1}{c|}{x} &
   &
   &
   &
   &
   &
   &
   \\ \hline
\begin{tabular}[c]{@{}l@{}}Amplified\\Movements\end{tabular} &
   &
   &
   &
   &
   &
   &
   &
   &
  \multicolumn{1}{c|}{x} &
   \\ \hline
Blur &
  \multicolumn{1}{c|}{x} &
  \multicolumn{1}{c|}{x} &
  \multicolumn{1}{c|}{x} &
  \multicolumn{1}{c|}{x} &
   &
   &
   &
   &
  \multicolumn{1}{c|}{x} &
   \\ \hline
Interval &
   &
   &
   &
   &
   &
  \multicolumn{1}{c|}{x} &
   &
   &
   &
   \\ \hline
\begin{tabular}[c]{@{}l@{}}Physiological Signals \\ Observation\end{tabular} &
   &
   &
   &
   &
   &
   &
   &
   &
   &
  \multicolumn{1}{c|}{x} \\ \hline
\end{tabular}
\end{table}

%The established attributes for the development stage of the experiment follow the observations, studies, and reports of the authors previously cited in this work.  We define pre-eliminated dataset with 34 attributes, divided into four groups, which are:

Our feature set is composed of 34 attributes, divided into four groups:

\begin{itemize}
\item \textbf{User Profile Data}: The profile data was selected based on the literature and also on the experience acquired during pilot tests of this work, we gathered this data through our Cybersickness Profile Questionnaire (CSPQ). The CSPQ contains questions such as gender, age, level of experience with VR, existence of any pre-symptoms and flicker sensibility. We also include information about whether the individual wears glasses, has any vision impairments, information regarding the posture during the experiment (standing or sitting), and which eye is the dominant eye.

\item \textbf{Questionnaire Data}: Information filled in by the user about discomfort symptoms before and after the experiment. The symptoms listed are from the VRSQ (Virtual Reality Sickness Questionnaire) ~\cite{kim2018virtualVRSQ}, which is a modified version of Kennedy's traditional SSQ (Simulator Sickness Questionnaire) to address environments virtual reality with HMDs specifically.

\item \textbf{Game data}: Information such as (1) time stamp, (2) speed, (3) acceleration, (4) player rotation axis, (5) player position, (6) region of interest, (7) size of the FOV, (8) frame rate and (9) discomfort level, which is the class and is reported by the user at any time during the gameplay experience.

\item \textbf{Game configuration data}: Boolean information towards the (1) existence of static resting frames, (2) existence of haptic response, (3) level of user control over the camera, (4) existence of depth of field simulation (DoF) and (5) whether the game primary camera moves automatically (without user intervention). These data were selected based on the list of strategies shown in Table ~\ref{tableStrategies}.

\end{itemize}

Two games were created for this work: a race and a flight game. The games (made with Unity 3D) are part of the complete software solution for this work. We collected data from a total 37 participants (9 female and 28 male) with ages ranging between 18 and 60. The participants were able to quit the experiment whenever they wanted. Each individual was asked to:

\begin{itemize}
    \item Task 1 - Fill in the profile questionnaire - CSPQ 
    \item Task 2 - Fill in the virtual reality sickness questionnaire VRSQ (Q1) 
    \item Task 3 - Participate in one of the VR games for up to 5 minutes (if possible).
    \item Task 4 - Fill in the virtual reality sickness questionnaire VRSQ (Q2).
\end{itemize}

\begin{figure}
\centering
\includegraphics[width=0.9\linewidth]{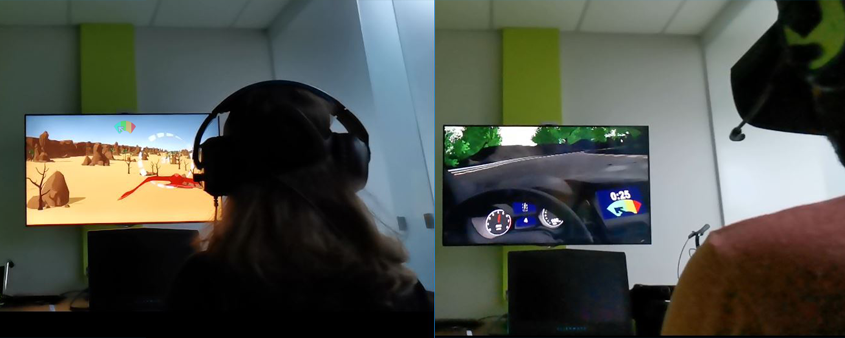}
\caption{The two games that were developed for this work (flight game at the left and race game at the right).}
\label{fig:userPlaying}
\end{figure}

Both games (racing and flight) try to force the participant to perform habitual VR game movements, such as rotation, translation, and perform acceleration changes. These movements can cause one or more symptoms of discomfort. Figure \ref{fig:userPlaying} shows two players using HDM playing both games (flight at left and race at right). %With the graphical visualizer application, it was possible to view a heatmap representation of accumulated CS-related by individuals (illustrated in Figure \ref{fig:discomfortAnalizer}).

Once the data is collected, machine learning algorithms were trained to classify the CS. Each of the inputs stored by the experiment contains a rating given by users related to discomfort (from 0 to 3, where 0 is none and 3 is severe) during gameplay.

%Before building and using the models, we performed an analysis with different classification algorithms in the WEKA \cite{hall2009weka} software.
%We considered different scenarios for training (binary and quarterly classification), in 3 subsets of the main dataset, which are:

To further analyse the collected data, we seperate the experiments in three main scenarios: A, B and C. In what follows, these scenarios are described:

\begin{itemize}
\item Scenario A - Classification consisting of data from the racing game (3993 samples).
\item Scenario B - Classification consisting of data from the flight game (5397 samples).
\item Scenario C - Classification using data from both scenarios together -A and B- (9390 samples).
\end {itemize}

\begin{figure}
\centering
\includegraphics[width=0.9\linewidth]{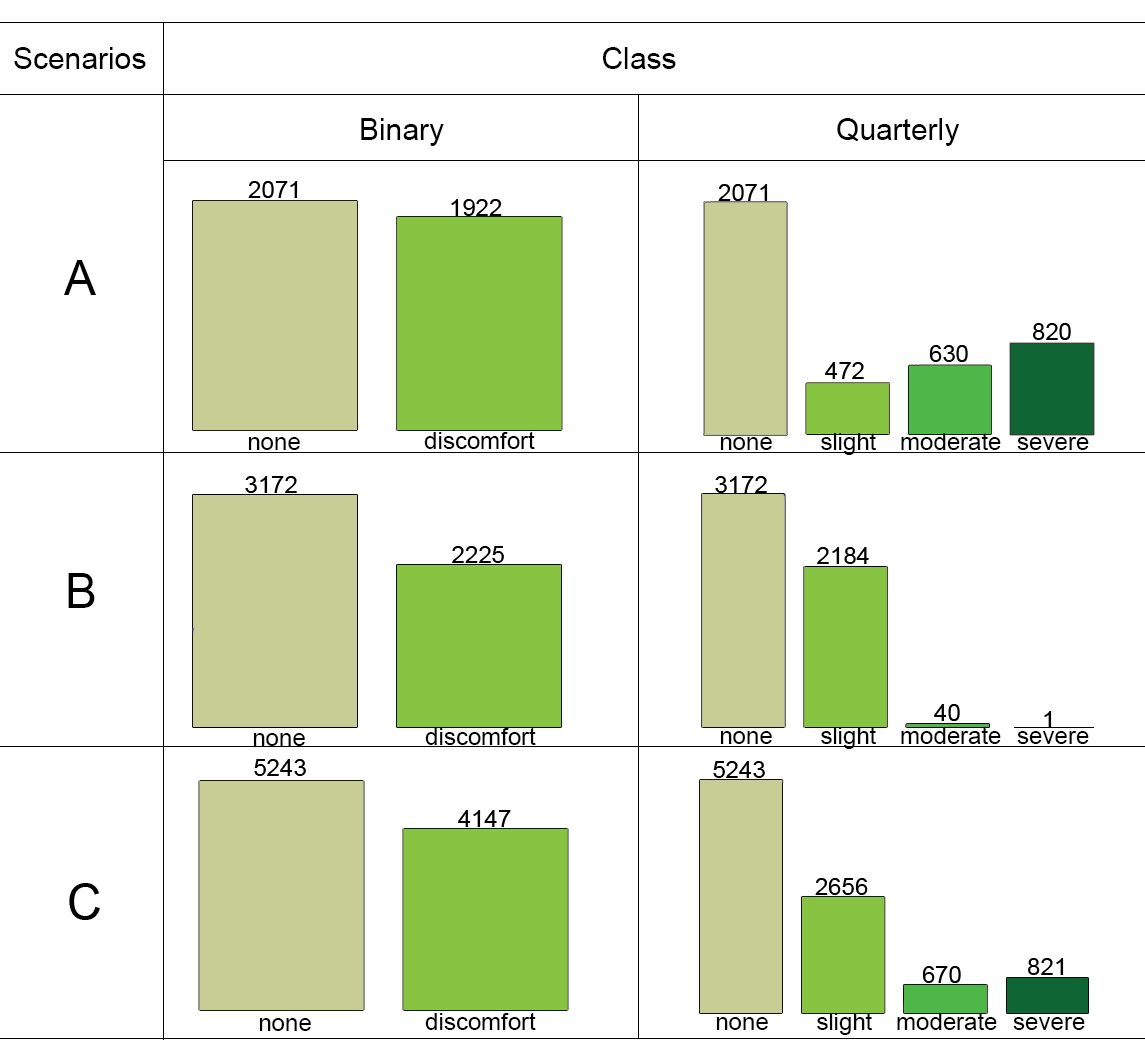}
\caption{Binary and quarterly class samples distribution in scenarios A,B and, C.}
\label{fig:DatabaseClass}
\end{figure}

The following classifiers were evaluated: BF Tree, CDT, Decision Strump, ForestPA, FT, Hoeffding, J48, J48 Graft, JCHAID Star, LAD Tree, LMT, Nb Tree, Random Forest, Random Tree, Rep Tree, and Simple Cart. Experiments were run using a 10-fold cross-validation in all scenarios. We also separated scenarios A, B and C into two new groups. The first group is a binary classification (0-none or 1-discomfort, which includes from slight to severe classes). The second group is a quaterly classification containing all four classes (none, slight, moderate and severe). The distribution of classes can be seen in Figure \ref{fig:DatabaseClass}. 

%In order to validate the classifications, we use accuracy and Kappa index value. The Kappa index was used for the analysis and validation of the classifiers because it is part of Weka's tools, and it represents an equivalent index value for all tested classifiers.

\section{Results}

\begin{figure}
\centering
\includegraphics[width=0.9\linewidth]{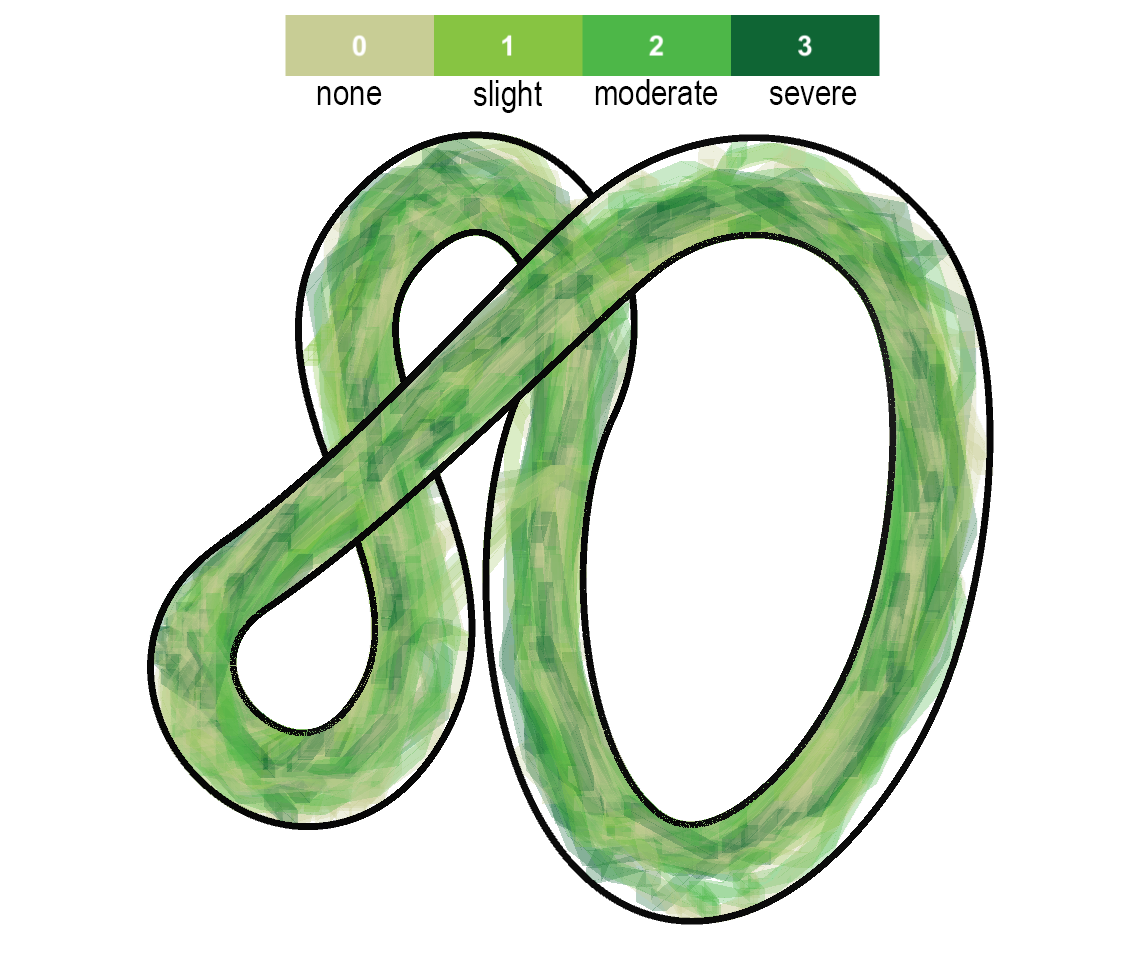}
\caption{Visualization of all moments where the participants of the elapsed game reported some of the levels of discomfort during the experiment. In the image, the intensity of discomfort reported by users varies from 0 (none) to 3 (severe) represented by the legend colors}
\label{fig:discomfortAnalizer}
\end{figure}

It was using the valid race game data with a total of 3993 samples. We observed that the occurrence of the discomfort reported by individuals occurs throughout the track (in Figure \ref{fig:discomfortAnalizer}). However, the discomfort levels in specifics regions of track have a more significant accumulation.

\begin{figure}
\centering
\includegraphics[width=0.9\linewidth]{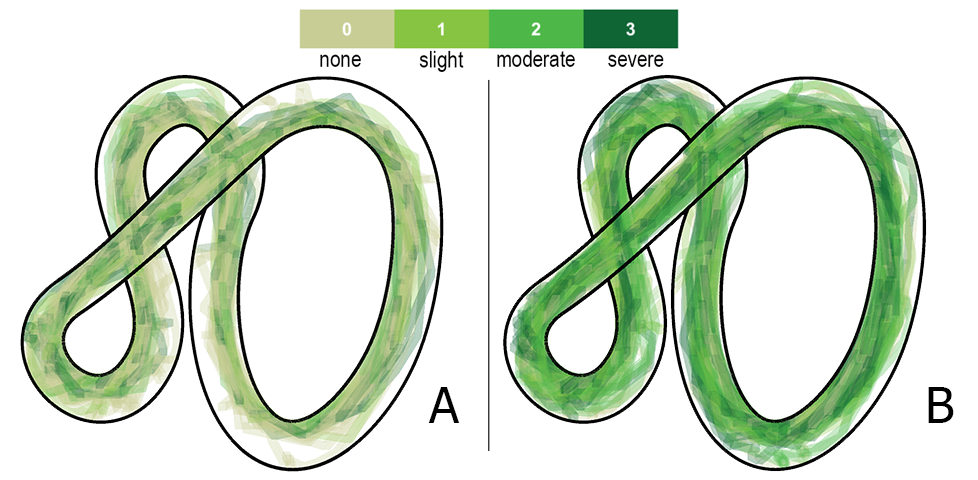} 
\caption{comparative of the discomfort reported by female (A) and male (B) participants. In the image, the intensities of discomfort reported by users vary from 0 (none) to 3 (severe) represented by the legend colors.}
\label{fig:AnalizerComparative}
\end{figure}

In a comparative sample of reports of discomfort between individuals of the female gender (7 females with 1772 samples) and male (8 males with 2221 samples), We observed that in an accumulated result, the male participants reported discomfort values greater than zero more often than individuals of the female gender (illustrated in Figure \ref{fig:AnalizerComparative}).

Biocca ~\cite {biocca1992will} and Kolasinski ~\cite {kolasinski1995simulator}, who report those female individuals are more susceptible to symptoms of MS. Despite being similar diseases, they have different environments and manifestations. Because of this, for this case, there is no way to say if there is a difference between genders for the manifestation of CS-based only on the works cited in this thesis. However, the results of this stage showed, in this specific testing stage, that the female audience reported less discomfort compared to the male audience.

%\section{Machine Learning Analysis}

%We divide into three main sub-sections, which are the results obtained by the binary and quarterly classifications in this paper. Each of the classifications was performed with 3 different sub-sets representing one of the described training scenarios previously (A, B, and C). 
Table \ref{tab:binaryClass},\ref{tab:quartelyClass} shows the Accuracy and Kappa index for binary and quarterly classifications over scenarios A, B and C.

\subsection{Binary Classification}
As previously mentioned, we also merged the discomfort level into a single class in order to perform binary classifications, which are usually stronger than non-binary ones. Discomfort values that were previously represented as slight, moderate, and severe are represented as discomfort.

In Scenario A, B, and C, the Random Forest classifier proved to be the best with an accuracy of 94.0\%, 99.0\%, and 96.6\% for the binary case.

% \usepackage{booktabs}
\begin{table}[]
\centering
\caption{Binary Classification}
\label{tab:binaryClass}
\begin{tabular}{@{}lllllll@{}}
\toprule
\multicolumn{7}{c}{Binary Classification}                            \\ \midrule
\multicolumn{1}{c|}{Scenarios} &
  \multicolumn{2}{c|}{A} &
  \multicolumn{2}{c|}{B} &
  \multicolumn{2}{c}{C} \\ \midrule
\multicolumn{1}{c|}{Classificator} &
  \multicolumn{1}{c|}{ACC} &
  \multicolumn{1}{c|}{KPP} &
  \multicolumn{1}{c|}{ACC} &
  \multicolumn{1}{c|}{KPP} &
  \multicolumn{1}{c|}{ACC} &
  \multicolumn{1}{c}{KPP} \\ \midrule
BFTree         & 91.8\% & 0.8364 & 96.8\% & 0.9353 & 93.0\% & 0.8584 \\
CDT            & 89.7\% & 0.794  & 96.8\% & 0.9346 & 92.3\% & 0.8457 \\
DecisionStrump & 62.0\% & 0.2574 & 71.9\% & 0.4299 & 61.6\% & 0.2713 \\
ForestPA       & 91.5\% & 0.8297 & 97.8\% & 0.9563 & 95.5\% & 0.909  \\
FT             & 87.0\% & 0.7395 & 94.2\% & 0.881  & 90.9\% & 0.8154 \\
Hoeffding      & 69.9\% & 0.393  & 78.1\% & 0.5363 & 71.9\% & 0.4251 \\
J48            & 92.4\% & 0.8495 & 97.9\% & 0.9576 & 95.2\% & 0.9036 \\
J48Graft       & 92.6\% & 0.853  & 97.9\% & 0.9575 & 95.2\% & 0.9036 \\
JCHAIDStar     & 89.8\% & 0.7968 & 92.8\% & 0.894  & 91.6\% & 0.8582 \\
LADTree        & 78.9\% & 0.5812 & 88.9\% & 0.7722 & 74.1\% & 0.4829 \\
LMT            & 93.0\% & 0.86   & 98.1\% & 0.961  & 95.5\% & 0.9088 \\
NbTree         & 88.1\% & 0.7624 & 98.6\% & 0.9728 & 95.0\% & 0.8993 \\
RandomForest   & 94.0\% & 0.8805 & 99.0\% & 0.9801 & 96.6\% & 0.9323 \\
RandomTree     & 89.2\% & 0.7838 & 96.6\% & 0.93   & 92.2\% & 0.8421 \\
RepTree        & 90.7\% & 0.8147 & 96.9\% & 0.9368 & 93.0\% & 0.8595 \\
SimpleCart     & 92.2\% & 0.8455 & 97.2\% & 0.9441 & 93.4\% & 0.8672 \\ \bottomrule
\end{tabular}
\end{table}

\subsection{Quarterly Classification}
For the quarterly classification, the dataset was kept unchanged. This set of experiments contain 4 classes: none, slight, moderate and severe. In Scenario A, the classifier LMT (Logistic Model Trees) achieved the best result, which reached the accuracy of 92.4\%. In the Scenario B, the best result was obtained with Random Forest (98.9\%). For Scenario C, Random Forest also obtained the best classification accuracy (95.4\%).

% \usepackage{booktabs}
\begin{table}[]
\centering
\caption{Quarterly Classification}
\label{tab:quartelyClass}
\begin{tabular}{@{}lllllll@{}}
\toprule
\multicolumn{7}{c}{Quarterly Classification}                          \\ \midrule
\multicolumn{1}{c|}{Scenarios} &
  \multicolumn{2}{c|}{A} &
  \multicolumn{2}{c|}{B} &
  \multicolumn{2}{c}{C} \\ \midrule
\multicolumn{1}{c|}{Classificator} &
  \multicolumn{1}{c|}{ACC} &
  \multicolumn{1}{c|}{KPP} &
  \multicolumn{1}{c|}{ACC} &
  \multicolumn{1}{c|}{KPP} &
  \multicolumn{1}{c|}{ACC} &
  \multicolumn{1}{c}{KPP} \\ \midrule
BFTree         & 88.8\% & 0.827  & 97.1\% & 0.9417 & 93.0\% & 0.8821 \\
CDT            & 86.3\% & 0.7854 & 96.5\% & 0.93   & 92.3\% & 0.8698 \\
DecisionStrump & 51.0\% & 0      & 71.6\% & 0.4294 & 55.8\% & 0      \\
ForestPA       & 87.6\% & 0.8017 & 97.4\% & 0.947  & 94.2\% & 0.9015 \\
FT             & 83.1\% & 0.7375 & 93.7\% & 0.8731 & 89.4\% & 0.8227 \\
Hoeffding      & 52.5\% & 0.0401 & 73.8\% & 0.4882 & 55.8\% & 0      \\
J48            & 90.7\% & 0.8566 & 97.8\% & 0.9569 & 94.8\% & 0.9139 \\
J48Graft       & 90.9\% & 0.8598 & 97.7\% & 0.9547 & 95.0\% & 0.9162 \\
JCHAIDStar     & 77.7\% & 0.6513 & 92.3\% & 0.8793 & 86.0\% & 0.7824 \\
LADTree        & 68.5\% & 0.4996 & 87.0\% & 0.7333 & 72.1\% & 0.4694 \\
LMT            & 92.4\% & 0.8832 & 97.8\% & 0.9566 & 95.5\% & 0.9249 \\
NbTree         & 88.7\% & 0.8246 & 98.7\% & 0.9747 & 94.4\% & 0.9052 \\
RandomForest   & 92.2\% & 0.8782 & 98.9\% & 0.9792 & 95.4\% & 0.9221 \\
RandomTree     & 85.1\% & 0.7709 & 96.8\% & 0.9355 & 89.5\% & 0.8243 \\
RepTree        & 87.0\% & 0.7962 & 96.7\% & 0.9328 & 92.6\% & 0.8755 \\
SimpleCart     & 88.9\% & 0.8288 & 97.3\% & 0.9464 & 93.0\% & 0.8821 \\ \bottomrule
\end{tabular}
\end{table}

\subsubsection{Attribute Selection}
%se quiser pode botar uma referencia do weka aqui dps que menciono weka
We used the Classifier Attribute Evaluator (using full set training with "leave one attribute out" method) in the Weka machine learning framework to generate a ranking of all attributes using the best classifier of the previous experiments (Tables \ref{tab:binaryClass} and \ref{tab:quartelyClass}). This algorithm removes attributes from the dataset and evaluates how its removal impacts on the performance of the classification. After all attributes are evaluated, they are ranked in terms of impact. The best ones stay at the top of the ranking.

For binary classifications and scenario A (racing game), the 5 most relevant attributes were: Time Stamp, Age, Gender, Rotation on the z axis (CameraRotationZ) and Player Speed. For Scenario B (flight game), the attribute ranking was given as follows: Age, Position on the z axis (PlayerPositionZ), Experience, Vision Impairment and Rotation on the z axis (CameraRotationZ).

These attributes ranked at the first 5 positions for the different scenarios corroborates with the literature. For 2 out of 3 test scenarios, the time stamp was considered the most important for the classification of discomfort. Player Speed, Rotation (CameraRotationZ) as well as Gender and Age were also essential attributes.

In quarterly classifications, we obtained similar results. For Scenario A: Time Stamp, Age, Genre, Experience and Player Speed were the top 5. Scenario B, the most relevant attributes were: Age, Position on the z axis (PlayerPositionZ), Experience and Vision Impairments. In Scenario C: Time Stamp, Age, Gender, Rotation on the z axis (CameraRotationZ), Experience.

\section{Conclusion}
%In this work, we carried out an extensive investigation of the existing literature, where theories were studied, and the leading causes related to discomfort in virtual reality environments were listed. A methodology was developed for this thesis, separating the final objective into 3 stages: Investigation, Experiment and, CS Automatic prediction.%, CS automatic strategy suggestion.

%The investigation stage contributed to the realization of the association between causes and strategies related to CS. With this association, it is possible to suggest strategies in the future once the cause of the discomfort in the virtual environment is known.

This is the first work to use classifiers in order to predict discomfort over the track during the gameplay experience considering a broad feature set, which includes user personal data. We also publicly provide the user data .

We built a data visualization application and performed a first analysis of the collected data. It is possible to observe that in 3993 samples of the racing game, most occurrences of discomfort reported by the participants occurred near or during curves of the virtual track, which reinforces the association of CS to rotations. We also made an analysis between genders which showed that female individuals reported lower incidents of discomfort compared to male participants in our data. 

Subsequently, the analysis in terms of machine learning consisted of three scenarios: Scenario A (data from the racing game), Scenario B (data from the flying game) and Scenario C (data from both games). We performed supervised binary and quarterly classifications using 16 decision tree classifiers. Classifiers that resulted in the highest accuracy were: Random Forests (Random Forest) and LMT (Logist Model Trees). The best accuracy was 99.0\% and was obtained with the Random Forest classifier for scenario B (flight game) in the binary classification. 

An attribute selection was also performed in order to identify the most relevant attributes. For all scenarios it was observed that the most relevant attributes were the same, being: (time, position, and z-axis rotation) and profile attributes of the individual. These results corroborates the importance of attributes related to the individuals in the prediction of CS. This assessment reinforces the theories and hypothesis present in the literature known so far. Time is notably essential for the prediction of discomfort in virtual reality environments. 

\section{Next Steps}

To construct a complete methodology, that contemplates the CS prediction and the automatic suggestion of strategies to minimize the CS. It will be necessary to overcome the following challenges:

\begin{itemize}
\item Application and analysis of association rules in the data set of this thesis.
\item Flight game analysis and inclusion of graphical visualization of discomfort.
\item Realization of a new stage of experiment with individuals standing.
\item Training with recurrent neural network model (LSTM).
\item Investigation of alternatives with deep learning networks.
\item Research on how to carry out automatic strategy suggestions with the value resulting from discomfort in real-time.
\item Inclusion of new attributes (in a synthetic way), with the captured data, for example, distances between the player and the movement limits of the scenarios, total distance covered.
\item Implementation of the final game application in virtual reality with the prediction of the level of discomfort and strategy suggestion in real-time.
\item Weight modeling for the profile questionnaire (CSPQ).
\end{itemize}

\begin{figure}
\centering
\includegraphics[width=0.9\linewidth]{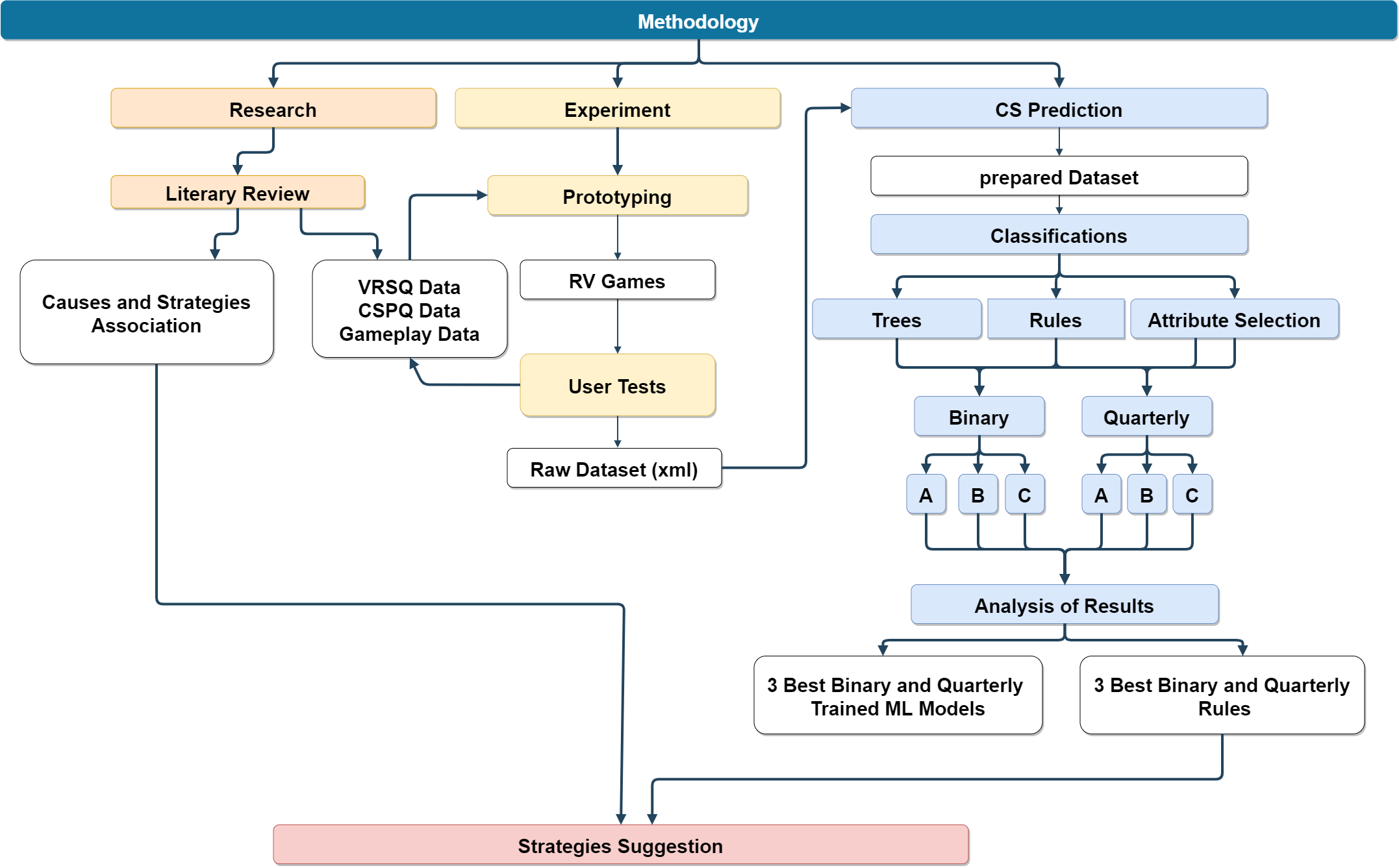}
\caption{Updated methodology diagram overview. At the present moment of this thesis, the stage of automatic strategy suggestion to minimize CS is listed as work in progress.}
\label{fig:Metodologia}
\end{figure}

As far as is known, most CS causes can have one or more minimization strategies. Based on this knowledge, the present methodology divides the problem of automatically suggesting in the following steps (Figure ~\ref {fig:Metodologia}):

\begin{itemize}
\item Research: Identification and association of causes and strategies
\item Experiments: Application development, tests with participants, data collection and pre-elimination analysis
\item CS Prediction (CS Prediction): classifiers, Training, binary classification, quaternary classification, attribute ranking, association rules.
\item Suggestion of CS minimization strategy (Strategies Suggestion).
\end{itemize}

\bibliographystyle{unsrt}  
\bibliography{references}  %%% Remove comment to use the external .bib file (using bibtex).

%%% Comment out this section when you bibliography{references} is enabled.

\end{document}